\documentclass[twocolumn,showpacs]{revtex4}
\usepackage{graphicx}
\usepackage{amsmath}
\usepackage{amsfonts}
\usepackage{amssymb}

\begin{document}
\title{The Entropy Bound for Local Quantum Field Theory}
\author{Yi-Xin Chen}
\email{yxchen@zimp.zju.edu.cn}
\author{Yong Xiao}
\email{yongxiao2008@live.cn}
\address{Zhejiang Institute of Modern Physics}
\address{Zhejiang University, Hangzhou 310027, China}

\begin{abstract}
We investigate the entropy bound for local quantum field theory in
this paper. Both the bosonic and fermionic fields confined to an
asymptotically flat spacetime are examined. By imposing the
non-gravitational collapse condition, we find both of them are
limited by the same entropy bound $A^{3/4}$, where $A$ is the
boundary area of the region where the quantum fields are contained
in. The gap between this entropy bound and the holographic entropy
has been verified.
\end{abstract}
 \maketitle

The counting of degrees of freedom of local quantum field theory
(LQFT) is a question of persistent interest. For example,
statistical mechanics tells that a thermal photon gas which is
described by LQFT has the entropy $S\sim l^{3}T^{3}$, when it is
confined to a box of size $l$. If the temperature $T$ could be an
arbitrarily chosen parameter, one finds the system has an entropy
proportional to the volume $l^{3}$. However, we know that this
temperature has to be limited by $E\sim l^{3}T^{4}\leqslant
E_{bh}\sim l$, or else the system will undergo collapse to form a
black hole. Substituting this physical limitation into the entropy
formula, one easily finds the entropy bound $S_{\max}\sim
l^{3/2}\sim A^{3/4}$, where $A$ is the boundary area of the
system. The derivation above is firstly given by 't Hooft in
\cite{hooft}. The entropy bound $A^{3/4}$ for LQFT in the absence
of black holes is also exemplified by other authors
\cite{cohen,hsu1,hsu2,hsu3,jack}.

On the other hand, there are still controversies around this
topic. Starting from a bosonic field model and imposing the
gravitational stability condition, a holographic entropy bound
which is proportional to the boundary area $A$ of the system is
derived in \cite{y}. This is a physically unaccepted outcome, for
that the holographic entropy bound $\frac{A}{4G}$ is laid by the
entropy of black hole \cite{bekenstein,hawking} the successful
counting of which in the context of string theory
\cite{vafa,witten} contains many states which are not describable
in conventional quantum field theory. It is conceivable that LQFT
should conform to a more stricter entropy bound rather than the
holographic entropy bound. Then it is worthy to note that Aste
\cite{aste1} has found a subtle mistake which is vital in the
derivation of \cite{y}, making the results there controversial
\cite{note1}.

In this paper we should devote ourselves to further verifying that
the entropy bound for LQFT is $A^{3/4}$ rather than $A$, since at
present there are seldom formal and general proofs of this fact
independent of the simple qualitative analysis above given by 't
Hooft. To do this, we shall revisit the model presented in
\cite{y} and developed by \cite{aste2}. This model originated by
Yurtsever \cite{y} is valuable in that it gives a very direct way
of counting degree of freedoms or the dimension of the Fock space
of a LQFT system. But notice that the imperfect considerations and
results in \cite{y,aste2} have been rectified here. The entropy
bound for both bosonic and fermionic fields are examined and found
to be $A^{3/4}$. We find that when the systems are far from
forming a black hole, surely a fermionic system has a much lower
entropy bound scaling than a bosonic system, as one may expect. We
shall also discuss the physical implications of the entropy gap
between LQFT and holographic theories, from $A^{3/4}$ to $A$. For
simplicity, we have set $G,\hbar,c,k_{B}=1$ in this paper. By
adding Planck length $l_{p}$ and Planck mass $m_{p}$ to the
expressions, one can readily get the right magnitudes. To make the
scaling behavior clear, the trivial constant coefficients will be
omitted in the calculations also.

We consider a system with massless scalar fields confined to a
3-dimensional spacelike cube of size $l$. Imposing periodic
boundary conditions, the particle's momentum will be quantized.
The elementary energy unit is equal to $\frac{\pi}{l}$, the
infrared (IR) energy cutoff of the system. Then any quantized wave
vector $\vec{k}$ can be labelled by three non-negative integers
$m_{x}$, $m_{y}$, $m_{z}$, that is $\left( k_{x}
,k_{y},k_{z}\right)=\frac{\pi}{l}\left( m_{x},m_{y},m_{z}\right)$.
The total number of these quantized modes is
\begin{align}
N=\sum\limits_{\vec{k}}1\sim\int d^{3}kd^{3}x=l^{3}\int
_{0}^{\Lambda}w^{2}dw=l^{3}\Lambda^{3},\label{1}
\end{align}
where $\Lambda$ is defined to be the ultraviolet (UV) energy
cutoff of LQFT. A sufficiently large volume is implied, thus the
summation over the discrete modes above can be replaced by the
corresponding integral. Due to Eq.\eqref{1}, the quantized wave
vector $\vec{k}$ can be one-to-one labelled by a character $i$
with $i\in\left[ 1,N\right]$. The corresponding energy of the mode
is $w_{i}=\left\vert \vec{k}\right\vert =\frac{\pi
}{l}\sqrt{m_{x}^{2}+m_{y}^{2}+m_{z}^{2}}$.

When the massless scalar fields obey the Bose-Einstein statistics,
we can construct the Fock states by assigning occupying number
$n_{i}$ to these $N$ different modes, which is
\begin{align}
\mid\Psi>=\mid n_{1},n_{2},\cdots n_{N}>\text{,
}n_{i}\in\mathbb{N}\text{, }i\in\left[  1,N\right]. \label{2}
\end{align}
Each different set of occupancy $\{n_{i}\}$ determines an
independent basis of the Hilbert space of the system. Now we
impose the non-gravitational collapse requirement which states
that the quantum states with energy more than the mass of a black
hole of the same size is unstable and thus should be excluded from
the physically permitted Hilbert space. It implies that
\begin{align}
E_{\Psi }=\sum_{i=1}^{N}n_{i}w_{i}\leqslant E_{bh}\sim l,
\label{3}
\end{align}
where $E_{bh}$ is the energy of the black hole with Schwarzschild
radius $r_{s}=l/2$. The number of solutions or occupancies
$\left\{n_{i}\right\}$ satisfying the requirement Eq.\eqref{3}
gives the dimension of the physically permitted Hilbert space
$W\equiv\dim\mathbb{H}$.

The entropy associated with the system is
$S=-\sum\limits_{j=1}^{W}\rho_{j}\ln\rho_{j}$ \cite{y,bekenstein},
where $\rho_{j}$ is the possibility distribution on the Hilbert
state basis. Obviously the maximum value of the expression can be
realized by a uniform distribution $\rho_{j}=\frac{1}{W}$. The
corresponding entropy is
\begin{align}
S_{\max}=-\sum\limits_{j=1}^{W}\frac{1}{W}\ln\frac{1}{W}=\ln W.
\label{4}
\end{align}
To determine the maximum entropy of the system, we have to count
out the dimension the Hilbert space, that is the number of
admissible solutions $\{n_{i}\}$ of Eq.\eqref{3}. This corresponds
to the knapsack or counting lattice points problem in mathematics
\cite{latt1,latt2,latt3}. That is, when adhering $\{n_{i}\}$ to a
Cartesian coordinate system $\{x_{i}\}$, the question refers to
the counting of lattice points (points with integer coordinates)
contained within the convex polytopes determined by
$\sum_{i=1}^{N}x_{i}w_{i}\leqslant E_{bh}$, $x_{i}\geqslant 0$,
with right-angle side lengths
\begin{align}
l_{i}=\frac{E_{bh}}{w_{i}}\text{, where }i\in\left[  1,N\right],
\label{5}
\end{align}
It is interesting but difficult to find an exact solution to this
question. However, the cases we refer to have $l_{i}\gg1$. Thus we
could use the volume of the corresponding polytopes to
approximately evaluate the number of lattice points within them
\cite{latt1,note2}.

To make the analysis explicit, we start from the counting of
quantum states with two different modes simultaneous excitated,
namely we require $n_{i_{1}},n_{i_{2}}\neq 0$, $n_{k}=0$ where
$k\neq i_{1},i_{2}$. Thus Eq.\eqref{3} reduces to
\begin{align}
n_{i_{1}}w_{i_{1}}+n_{i_{2}}w_{i_{2}}\leqslant E_{bh}.\label{6}
\end{align}
Consider a $2$-dimensional polytope $\mathcal{P}^{2}$ with
right-angle size lengths $\frac{E_{bh}}{w_{i_{1}}}$ and
$\frac{E_{bh}}{w_{i_{2}}}$. There is an one-to-one correspondence
of the solutions $\{n_{i_{1}},n_{i_{2}}\}$ of Eq.\eqref{6} with
the lattice points in $\mathcal{P}^{2}$. Since each integer cell
occupies one unit volume in the polytope, we can approximately
evaluate the number of lattice points within $\mathcal{P}^{2}$ by
its 2-dimensional volume $Vol\left(  \mathcal{P}^{2}\right)
=\frac{1}{2!}l_{i_{1} }l_{i_{2}}$. Summing over the choices of
$i_{1}$, $i_{2}$, the total number of states with two modes
simultaneously excitated is
$\frac{1}{2!}\sum\limits_{i_{1}<i_{2}}^{N}l_{i_{1}}l_{i_{2}}$.

Similarly, we estimate the number of quantum states with $m$ modes
simultaneously excitated by virtue of the volume of the related
polytopes $\mathcal{P}^{m}$ as
$Vol\left(\mathcal{P}\right)=\frac{1}{m!}S_{m}$, where
\begin{align}
S_{m}\equiv\sum\limits_{i_{1}<i_{2}<\cdots
i_{m}}^{N}l_{i_{1}}l_{i_{2}}\cdots
l_{i_{m}}<\frac{1}{m!}\left(\sum_{i=1}^{N}l_{i}\right)^{m},\label{7}\\
z\equiv\sum_{i=1}^{N}l_{i}\sim l^{3}
\int_{0}^{\Lambda}\frac{E_{bh}}{w}\cdot w^{2}dw\propto
E_{bh}l^{3}\Lambda^{2}. \label{8}
\end{align}
In belief, this entire procedure corresponds to evaluating the
solutions of Eq.\eqref{3} by counting the lattice points contained
within $\mathcal{P}^{N}$ and its lower dimensional surfaces
$\mathcal{P}^{m}$. In this way we get the dimension of the Hilbert
space
\begin{align}
W=\sum\limits_{m=0}^{N}\frac{1}{m!}S_{m}<\sum\limits_{m=0}^{N}\frac{1}{\left(
m!\right) ^{2}}z^{m}.\label{9}
\end{align}
The summation in Eq.\eqref{9} is up to $S_{N}$. It means that the
states with $N$ modes simultaneously excitated could contribute to
the counting of the physical Hilbert space. To insure this, we
consider a $N$ particle state with one particle occupying one
mode. This state is the lowest energy state with $N$ modes
simultaneously excitated and at least should satisfy the
gravitational stability condition Eq.\eqref{3}. It gives
\begin{align}
l^{3}\int_{0}^{\Lambda}w\cdot w^{2}dw\sim
l^{3}\Lambda^{4}\leqslant E_{bh}\sim l.\label{10}
\end{align}
It is not other but the well-known UV-IR relation for LQFT first
established by Cohen et al \cite{cohen} to exclude all
non-gravitational stability states that lie within their
Schwarzschild radius. For convenience, it is usually written as
${\Lambda \leqslant l}^{-1/2}$.

Combining the above results with the asymptotic form of the Bessel
function, we have
\begin{align}
W<\sum\limits_{m=0}^{N}\frac{1}{\left(  m!\right)
^{2}}z^{m}\leqslant \sum\limits_{m=0}^{\infty}\frac{1}{\left(
m!\right)  ^{2}}z^{m}\sim \exp\left(  2\sqrt{z}\right). \label{11}
\end{align}
Notice that the second inequality can be saturated at $\Lambda
=\Lambda _{\max }=l^{-1/2}$. The reason is as below. Since both
$z$ and $N$ are functions of $\Lambda $, we find $\sqrt{z}\sim
l^{3/2}\sim N$ when $\Lambda =l^{-1/2}$. In this case there is
$\frac{1}{(m!)^{2}}z^{m}\sim \left( \frac{N}{m}\right) ^{2m}\sim
0$ ($m!\sim m^{m}$, $m>N$) which makes the use of
$\sum\limits_{m=0}^{N}\sim \sum\limits_{m=0}^{\infty}$ reasonable.

From Eq.\eqref{4} and Eq.\eqref{11}, the maximum entropy is
evaluated as \cite{note3}
\begin{align}
S_{\max}=\ln W<\left(E_{bh}l^{3}\Lambda^{2}\right)^{1/2}\sim l^{2}
\Lambda. \label{12}
\end{align}
If the UV cutoff $\Lambda$ is taken as an arbitrary constant, one
will find a holographic entropy bound proportional to the area
$A\sim l^{2}$ of the system. But actually taking $\Lambda$ always
as constant will cause troubles, such as the problem presented in
\cite{aste1} aiming to invalid the results of \cite{y}. By
contrast, we have pointed out that $\Lambda$ should limited by the
UV-IR relation $\Lambda\leqslant l^{-\frac{1}{2}}$ for LQFT.
Substituting it into Eq.\eqref{12}, we get the right entropy bound
$S_{\max}\leqslant A^{3/4}$.

Notice that the UV-IR relation $\Lambda\leqslant l^{-\frac{1}{2}}$
plays an essential role in the evaluation of the entropy bound for
LQFT. The validity of it for LQFT has been argued in the
literature for various applications. In \cite{cohen}, it has been
pointed out that this UV-IR relation is necessary in order to make
sure LQFT could be a good effective description of the nature.
Furthermore, it was also illustrated in \cite{hsu3,us} that when
$\Lambda>l^{-\frac{1}{2}}$, the gravitational corrections to the
energy of the system will be too large and lead to gravitational
collapse, which makes a LQFT description invalid. This UV-IR
relation even has found applications in cosmology to establish
dark energy models \cite{hsu3,us,limiao}. As far as the present
model is concerned, one can check the self-consistency of the use
of the UV-IR relation in our calculation. When Eq.\eqref{10} is
saturated, it implies that only one state with $N$ modes
simultaneously excitated could exist. At the same time, the volume
counting method gives $\frac{1}{(m!)^{2}}z^{m}\sim \left(
\frac{N}{m}\right) ^{2m}\sim 0$ with $m>N$, which indicates that
no states will be counted in the evaluation Eq.\eqref{11} when
$m>N$. This consistency is more than could be expected.

The analysis of entropy bound for fermionic fields is
straightforward and will coincides with that given by Cohen et al
\cite{cohen} for fermionic and compact bosonic fields. From the
Fermi-Dirac statistics, occupancy number for any mode is simply
$0$ or $1$. The UV-IR relation Eq.\eqref{10} insures that the
state with all the $N=l^{3}\Lambda^{3}$ modes being occupied
satisfies the gravitational stability requirement. It has been the
maximum energy state in the fermionic system. Thus all the
fermionic states can really contribute to the dimension of the
Hilbert space, which gives $W=2^{l^{3}\Lambda^{3}}$. So the
maximum entropy is $S_{\max}=\ln W\sim l^{3}\Lambda^{3}$. When
$\Lambda=\Lambda_{\max}=l^{-1/2}$, which means the energy of the
system comes close to the critical energy to form a black hole,
the fermionic system gives the same entropy bound $A^{3/4}$ as
that of bosonic systems. However, when $\Lambda\ll l^{-1/2}$,
there is $l^{3}\Lambda^{3} \ll l^{2}\Lambda$. It implies that when
the energy of the system is far from the formation of a black
hole, surely fermionic fields will have a much lower entropy bound
scaling than that of bosonic fields.

The generalizations to $D$-dimensional space-time and quantized
fields with polarization or higher spins are straightforward.
Following the steps above, one can readily find the entropy bound
scaling as $\left( E_{bh}l^{D-1}\Lambda ^{D-2}\right) ^{1/2}$ for
bosonic fields and $\left(l\Lambda\right)^{D-1}$ for fermionic
fields, along with the UV-IR relation $\Lambda\leqslant
L^{-\frac{2}{D}}$. Remembering that $A=l^{D-2}$ and $E_{bh}\sim
l^{D-3}$, the entropy bound $A^{\left( D-1\right)/D}$ is retrieved
for all bosonic and fermionic fields in $D$ dimensions.

We have imposed an energy cutoff $\Lambda$ in our derivation. The
modes with energy more than the UV cutoff have been excluded in
our consideration. But one may still argue that this cutoff can
only be justified in an average sense, and worry about the states
where some modes with momentum $k>\Lambda$ are populated but the
total energy is less than the size of the system $l$. Actually,
the introduction of UV cutoff is a useful technique in the
regularization of QFT. It doesn't deny minor fluctuations
deviating from the introduced macroscopic parameter. The point is
that these fluctuations will not cause considerable contributions
to the final results. Here we write a rough calculation to clarify
this issue. We start directly from Eq.\eqref{3} to evaluate the
number of physically permitted states, without an additional
assumption about the cutoff. We use $\Lambda_{m}$ as the highest
reachable energy with $m$ modes simultaneously excitated, which
can be approximated taken as $\frac{E_{bh}}{m}$. (Or else the
energy of the corresponding states will exceed the mass of a black
hole of the same size.) Similar steps to these from Eq.\eqref{7}
to Eq.\eqref{9} lead to
\begin{align}
W<\sum\limits_{m=0}^{\infty}\frac{1}{\left(  m!\right)
^{2}m^{2m}}\left( E_{bh}l\right)^{3m}.\label{13}
\end{align}
From Stirling's formula, the dominant contribution to $W$ comes
from $m_{0}\sim\left(  E_{bh}l\right)  ^{3/4}\sim A^{3/4}$. With
$W\sim e^{2m_{0}}$, the induced entropy bound is still $A^{3/4}$.
The states with more modes excitated are not permitted, since
$\frac{1}{\left(m!\right) ^{2}m^{2m}}\left(E_{bh}l\right)
^{3m}\sim0$ when $m>A^{3/4}$. \ Here
$\Lambda_{m_{0}}\sim\frac{E_{bh}}{m_{0}}\sim l^{-1/2}$. Thus the
UV-IR is still effective with a slightly different interpretation.
It is meaningful to note that the above estimation can also be
generalized to higher dimensions and lead to the entropy bound
$A^{\left(D-1\right)/D}$.

Now we turn to discuss the implications of our results. There is
surely an entropy gap between LQFT and some unknown holographic
theories, from $A^{3/4}$ to $A$ \cite{us}. Actually, it has been
pointed out in \cite{cohen} that the holographic theories should
have an UV-IR relation $\Lambda\leqslant l^{-\frac{1}{3}}$,
compared to $\Lambda\leqslant l^{-\frac{1}{2}}$ for the LQFT. The
higher cutoff energy implies that one could detect more finer
structures of the spacetime in holographic theories, since the UV
cutoff determines the minimal detectable lengths. Thus the
holographic theories would involve more degrees of freedom than
that in the LQFT, which will fill the entropy gap between them.
Nevertheless, the statistical principles of the holographic
theories and the microscopic origin of the holographic UV-IR
relation are still obscure. It inevitably involves the evaluation
of quantum-gravitational degrees of freedom. The efforts in this
direction will shed light on the understanding of the holographic
principle \cite{hooft,susskind,bousso}. It will also be
interesting if one can explicitly count out all the holographic
degrees of freedom similar to these we did for LQFT.

To support our arguments on the entropy gap, a definite example is
useful. (For details, see our recent paper \cite{us}.) For a
homogeneous and isotropic universe with FRW metric, the energy
contained within the apparent horizon is $U=\frac{r_{a}}{2}$ where
$r_{a}$ is the radius of the apparent horizon of the universe.
This energy is formally identical to the mass of a black hole of
the same size. Respecting the thermodynamics law $dU=TdS+pdV$, for
universes with $T\sim r_{a}^{-1}$ like the de Sitter universe, one
gets the holographic entropy $S\sim r_{a}^{2}=A$ where $A$ is the
area of the apparent horizon. By contrast, for universes dominated
by conventional fields, such as radiation, we have $T\sim
r_{a}^{-1/2}$. This different temperature behavior requires the
entropy scaling is $S\sim r_{a}^{3/2}=A^{3/4}$ to make sure that
$TdS$ is comparable to $dU=\frac{1}{2}dr_{a}$.

We have given a consistent derivation to the entropy bound for
LQFT in this paper. The right entropy bound $A^{\frac{3}{4}}$ is
obtained for both bosonic and fermionic systems. The necessity of
introducing the UV-IR relation is explained. The role and
implications of the UV-IR relation for LQFT and cosmology have
been partly investigated in previous works
\cite{cohen,us,limiao,michael} and are still worthy of further
study. The entropy gap between LQFT and holographic theories has
been verified. The implications of it should be clarified in
further studies \cite{hsu2,us}.

\section*{Acknowledgements} We would like to thank C. Cao for useful
discussions. The work is supported in part by the NNSF of China
Grant No. 90503009, No. 10775116, and 973 Program Grant No.
2005CB724508.


\begin{thebibliography}{99}
\bibitem{note2}Actually, due to the UV-IR relation Eq.\eqref{10},
for even the shortest edge $l_{N}$ of these polytopes, there is
$l_{N}=\frac{E_{bh}}{\Lambda }\geqslant N=l^{3}\Lambda ^{3}\gg 1$.
Thus we can apply the volume estimation of the number of lattice
points without doubt.
\bibitem{note3}In the derivation to Eq.\eqref{12}, we have not
use the key steps from Eq.[22] to Eq.[28] in \cite{y}, where
$S_{m}$ is expressed as the recursion formula
$S_{m}=\frac{1}{m}\sum\limits_{j=1}^{m}\left( -1\right)
^{j}P_{j}S_{m-j}$ and then extract $\frac {1}{m!}\left(
P_{1}\right)^{m}$ from it as the dominant term, therein $P_{1}\sim
E_{bh}l^{3}\Lambda^{2}$, $P_{j}\sim\left( E_{bh}l\right) ^{j}$
($4\leqslant j\leqslant N$). The reason is that those steps are
defective. To clarify possible misunderstandings, we make a
compact explanation here. Actually we find that the recursion
system is solvable and can be expressed as $S_{m}=\frac
{1}{m!}\sum\frac{m!}{a_{1}!a_{2}!\cdots a_{N}!}(P_{1})^{a_{1}}(-P_{2}%
/2)^{a_{2}}\cdots\left(\left(-1\right)^{N-1}P_{N}/N\right)
^{a_{N}}$, here the summation is over $a_{1}+2a_{2}+\cdots
Na_{N}=m$. Obviously the asymptotic expansion is
$\dfrac{1}{m!}[(P_{1})^{m}-\frac{1}{2}m^{2}P_{2}\left(
P_{1}\right)  ^{m-2}+\frac{1}{3}m^{3}P_{3}\left(  P_{1}\right)
^{m-3}+\cdots +(-1)^{m-1}(m-1)!P_{m}]$. When
$l^{2}\Lambda^{2}<m<N$, terms like $m^{j}P_{j}\left( P_{1}\right)
^{m-j}$ will overcome the term $\left( P_{1}\right)  ^{m}$. Thus
one cannot extract $\frac {1}{m!}\left( P_{1}\right)^{m}$ from the
recursion system as did in \cite{y} where the contributions from
$m^{j}$ have been incorrectly ignored. The struggle to take all
the other terms into account will also turn into a disaster.
Because of the presence of $(-1)^{j}$, these terms counteract with
each other, making one cannot even know the final corrections be
negative or positive. Thus we conclude that the expression of
$S_{m}$ as the recursion formula can not be even helpful in the
evaluation. By contrast, we adopt a more advisable derivation from
Eq.\eqref{7} to Eq.\eqref{12} and avoid those defects in \cite{y}.
\end{thebibliography}
\end{document}